\documentclass[preprint,12pt]{elsarticle} 

\usepackage{hyperref}
\usepackage{upgreek}
\usepackage{array,multirow}
\usepackage{url}
\usepackage{ulem}
\usepackage{xcolor}

\usepackage{amssymb}

\newcounter{bla}

\journal{Computer Physics Communications}

\begin{document}

\begin{frontmatter}



\title{\boldmath PARSIFAL: a toolkit for triple-GEM parametrized simulation}

\author[a,b]{A.~Amoroso}
\author[c]{R.~Baldini~Ferroli}
\author[d,e]{I.~Balossino}
\author[c]{M.~Bertani}
\author[d]{D.~Bettoni}
\author[a,b]{F.~Bianchi}
\author[a,b]{A.~Bortone}
\author[c]{A.~Calcaterra}
\author[c]{S.~Cerioni\fnref{cerioni}}
\fntext[cerioni]{Deceased}
\author[a]{W.~Cheng}
\author[d]{G.~Cibinetto}
\author[d]{A.~Cotta~Ramusino}
\author[a,b]{G.~Cotto}
\author[a]{F.~Cossio}
\author[a]{M.~Da~Rocha~Rolo}
\author[a,b]{F.~De~Mori}
\author[a,b]{M.~Destefanis}
\author[e]{J.~Dong}
\author[d,f]{F.~Evangelisti}
\author[d]{R.~Farinelli} 
\author[a]{L.~Fava}
\author[c]{G.~Felici}
\author[d,f]{I.~Garzia}
\author[c]{M.~Gatta}
\author[a]{G.~Giraudo}
\author[d,f]{S.~Gramigna}
\author[a,b]{M.~Greco}
\author[a,b]{L.~Lavezzi\corref{1}} \ead{lia.lavezzi@to.infn.it}
\author[a,b]{M.~Maggiora}
\author[d]{R.~Malaguti}
\author[g,h]{A.~Mangoni}
\author[a,b]{S.~Marcello}
\author[d]{M.~Melchiorri}
\author[d,e]{G.~Mezzadri}
\author[c]{E.~Pace}
\author[g,h]{S.~Pacetti}
\author[c]{P.~Patteri}
\author[a,b]{J.~Pellegrino \fnref{jacopo}}
\fntext[jacopo]{Currently at: right. based on science GmbH}
\author[a]{A.~Rivetti}
\author[d,f]{M.~Scodeggio}
\author[a,b]{S.~Sosio}
\author[a,b]{S.~Spataro}

\address[a]{INFN, Sezione di Torino, via P. Giuria 1, 10125 Torino, Italy}
\address[b]{Universit\`a di Torino, Dipartimento di Fisica, via P. Giuria 1, 10125 Torino, Italy}
\address[c]{INFN, Laboratori Nazionali di Frascati, via E. Fermi 40, 00044 Frascati (Roma), Italy}
\address[d]{INFN, Sezione di Ferrara, via G. Saragat 1, 44122 Ferrara, Italy}
\address[e]{Institute of High Energy Physics, Chinese Academy of Sciences, 19B YuquanLu, Beijing, 100049, People's Republic of China}
\address[f]{Universit\`a di Ferrara, Dipartimento di Fisica e Scienze della Terra, via G. Saragat 1, 44122 Ferrara, Italy}
\address[g]{INFN, Sezione di Perugia, via A. Pascoli, 06123 Perugia, Italy}
\address[h]{Universit\`a di Perugia, Dipartimento di Fisica e Geologia, via A. Pascoli, 06123 Perugia, Italy}

\cortext[1]{Corresponding author}

\begin{abstract}
PARSIFAL (PARametrized SImulation) is a fast and reliable software tool that reproduces the complete response of a triple-GEM detector to the passage of a charged particle, taking into account the main physical effects. Starting from the detector configuration and the particle information, PARSIFAL reproduces ionization, spatial and temporal diffusion, effect of magnetic field, if present, and GEM amplification to provide the dependable triple-GEM detector response. In the design and optimization stages of this kind of detectors, simulations play an important role. Accurate and robust software programs, such as GARFIELD++, can simulate the transport of electrons and ions in a gas medium and their interaction with the electric field, but they are CPU-time consuming. The necessity to reduce the processing time while maintaining the precision of a full simulation is the main driver of this work. For a given set of geometrical and electrical settings, GARFIELD++ is run once-and-for-all to provide the input parameters for PARSIFAL. Once PARSIFAL is initialized and run, it produces the detector output, including the signal induction and the output of the electronics. The results of the analysis of the simulated data obtained with PARSIFAL are compared with the results of the experimental data collected during a testbeam: some tuning factors are applied to the simulation to improve the agreement. This paper describes the structure of the code and the methodology used to match the output to the experimental data.
\end{abstract}


\end{frontmatter}



{\bf PROGRAM SUMMARY}

\begin{small}
\noindent
{\em Program Title:} PARSIFAL \\
{\em CPC Library link to program files:} (to be added by Technical Editor) \\
{\em Developer's repository link:} https://github.com/Hilldar/PARSIFAL (branch\\ triplegem) \\
{\em Code Ocean capsule:} (to be added by Technical Editor)\\
{\em Licensing provisions:} GPLv3  \\
{\em Programming language:} C++ \\
{\em Nature of problem:} Monte Carlo (MC) simulations are widely used in design and development of detectors for high energy physics as well as during data taking to understand how the detector geometry, acceptance, efficiency affect the experimental observations and hence infer the systematic effects. For triple-GEM detectors (one of the most used Micro Pattern Gaseous Detectors), the simulation of their response to the passage of particles is usually performed with GARFIELD++ (CERN) which provides a microscopic description of the signal creation, from the interaction of the particle with the gas to the induction of the signal on the anode. This detailed simulation is heavy and CPU-time consuming, hence it requires long computing periods to gain high statistics.\\
{\em Solution method:} PARSIFAL software provides the MC simulation of the response of a triple-GEM detector to the passage of a charged particle by splitting the simulation into four independent steps: ionization, electron drift in gas and magnetic field, avalanche formation in the multiplication stages, signal induction on the anode and response of the electronics (APV-25 chip). For each step, PARSIFAL samples the variables of interest from distributions, obtained by using a set of input parameters. These parameters are extracted from a simulation run only once with GARFIELD++. Since the sampling is much faster than a full GARFIELD++ simulation, this reduces the CPU-time to collect a sample with high statistics. The results extracted by the PARSIFAL simulation have been tuned with experimental data collected during a testbeam at CERN and show a satisfactory compatibility. 

\end{small}


\section{Introduction} \label{sec:intro}
\begin{figure}[b!]
\centering
\includegraphics[width=0.27\textwidth]{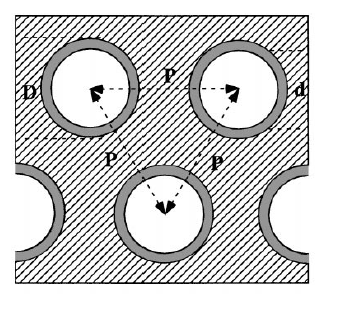} \qquad
\includegraphics[width=0.34\textwidth]{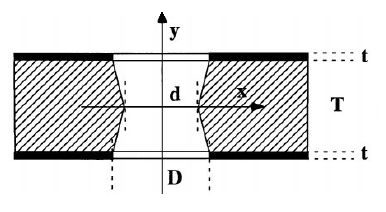} 
\qquad
\includegraphics[width=0.25\textwidth]{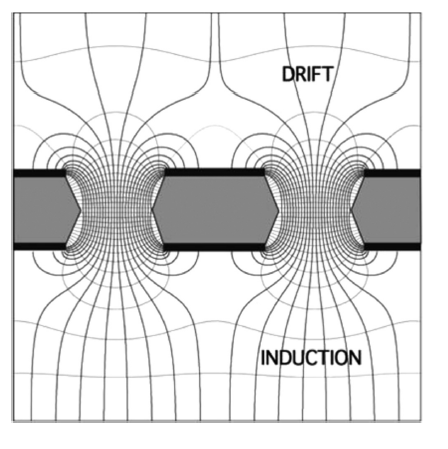} 
\caption{Scheme of a GEM foil from top ({\it left}) and side ({\it middle}): the holes are equidistant with a pitch $p = 140\,\upmu$m, they have a bi-conical shape with inner diameter $d = 50\,\upmu$m and outer diameter $D = 70\,\upmu$m. The copper thickness $t$ is $5~\upmu$m and the polyimide thickness $T$ is $50~\upmu$m \cite{bachmann2}. A high voltage difference is applied between the copper layers and the electric field lines generated inside the holes are shown in the {\it right} plot \cite{sauli2}.} \label{fig:gem}
\end{figure}

\begin{figure}[htb!]
\centering
\includegraphics[height=0.29\textwidth]{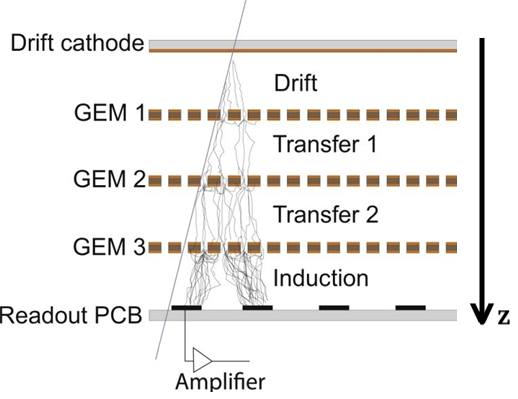}
\hspace{1cm}
\includegraphics[height=0.29\textwidth]{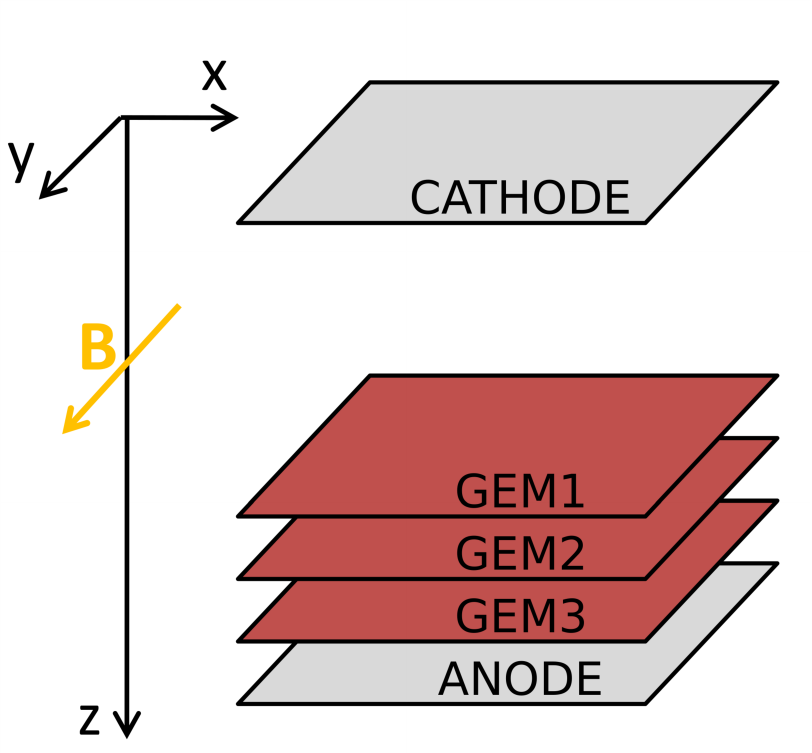} 
\caption{Schematic of a triple-GEM tracker showing an example of ionizing track and signal amplification ({\it left}) \cite{cmstdr}. The broken lines represent the electron drift paths. The number of lines increases after each stage of amplification up to the segmented anode, where the electrons are collected. Sketch of a tridimensional triple-GEM chamber, to illustrate the reference axes orientation assumed in this paper and the direction of the magnetic field (B), when it is applied ({\it right}).} \label{fig:triplegem}
\end{figure}

In 1997, F. Sauli \cite{sauli} introduced a new technique for signal amplification in gaseous detectors: the Gaseous Electron Multiplier (GEM). It consists of a $50\,\upmu$m thick polyimide foil, copper coated ($5\,\upmu$m) on both sides, with a high density pattern of holes with a diameter of $50\,\upmu$m and a pitch of $140\,\upmu$m (Figure~\ref{fig:gem}, {\it left, middle}). The application of an electric potential difference of a few hundreds of volts between the two copper sides produces an electric field of about $100$~kV/cm inside the holes. When an electron, generated by the ionization of the specific gas mixture, enters a hole, the intense field is sufficient to create an avalanche multiplication (Figure~\ref{fig:gem}, {\it right}). Typically, multiple stages of amplification are stacked to achieve the desired gain value, with a lower voltage applied to each stage. This reduces the probability of discharge compared to a single-stage GEM with the same amplification gain \cite{bachmann}. 
Figure~\ref{fig:triplegem}, {\it left}, shows a schematic drawing of a typical triple-GEM detector configuration, which consists of a cathode, three stages of GEM and an anode for the signal readout, segmented in strips or pads. 

A software tool that can describe the response of a detector system and compute its performance is of great importance in the design and optimization phase of the detector itself, as well as for a running experiment. The simulations	must reproduce the experimental measurements with an excellent level of agreement, so they can be used to predict the detector behavior with different settings. \\
The most widespread and robust existing software for gaseous detector simulation is GARFIELD++, which is defined by its authors in \cite{garfield0,garfield1,garfield2} as {\it``an object-oriented toolkit for the detailed simulation of particle detectors which use a gas mixture or a semiconductor material as sensitive medium''}. \\
It provides interfaces to additional packages, such as HEED \cite{heed}, for primary and secondary ionizations, and MAGBOLTZ \cite{magboltz}, to describe the electron diffusion effect and the drift in magnetic and electric fields. The input map of electric fields can be provided by external tools, {\it e.g.} ANSYS \cite{ansys}. The interaction between electrons and gas molecules is described at microscopic level. \\
A triple-GEM simulation performed by GARFIELD++, however, is CPU-time demanding. In fact, the entire path of the electron along the electric field lines, under the possible effect of the magnetic field, is simulated in steps of a few microns. In each step, GARFIELD++ simulates all the interactions of the electron with the gas mixture and the effect of the fields. This is done for each electron and the total number of electrons is about $10^5$-$10^6$. Such a level of detail is the reason why GARFIELD++ simulations are quite expensive in terms of CPU-time. Examples of triple-GEM simulations performed with GARFIELD++ can be found in the literature \cite{garf1,garf2,garf3}. For studies requiring high statistics, however, it is mandatory to reduce the time by several orders of magnitude. In the literature, there are already attempts to cope with the triple-GEM simulation by splitting it into independent tasks, each considering one physical process \cite{bonivento}. \\
The simulation of separate processes ({\it i.e.} diffusion and amplification) is much less CPU-time consuming compared to a single simulation. This modular approach is the first technique used to speed up the code. It allows easy modification of the detector configuration or incoming particle information and permits the extension of the code to other Multi Pattern Gaseous Detectors (MPGD). The second method to boost the simulation is to adopt a parametrized simulation, instead of a step-by-step one. The parameters of interest described in Section~\ref{sec:gsim} are extracted from a GARFIELD++ simulation, run once-and-for-all to initialize PARSIFAL. Once the configuration is set, PARSIFAL reproduces the detector output with reduced CPU-time and high statistics. \\
Such a capability is important both in $R\&D$ of the detector and after its installation inside an experimental setup, at data taking stage. In fact, during $R\&D$, a trustworthy simulation, validated with data in a specific range of settings, can be run with different settings to evaluate the detector performance. On the other hand, at experiment run-time, the geometrical and electrical settings of the detector remain quite stable, while different event topologies have to be simulated and compared with the experimental data. In this case, the simulation requires good reliability and can benefit from the high speed of the software.\\
PARSIFAL is a software capable of producing a comprehensive, reliable and fast simulation of GEM-based detectors, which follows the same approach as \cite{bonivento}, with some additional implementations and improvements. The code is written in C++, Object-Oriented and requires only ROOT framework \cite{root} for installation and compilation. It needs to be initialized with parameters that can be extracted from GARFIELD++, but it is not directly interfaced to this program, so it can run independently from it.
Algorithms for position reconstruction are also implemented in the software PARSIFAL: the reconstructed positions are used to evaluate the efficiency and spatial resolution of the simulated detector. \\
The paper is organized as follows: Section~\ref{sec:setup} describes the detector configuration implemented in the software, with geometrical and electrical characteristics matching those of the testbeam, to allow for a direct comparison of the simulated and experimental results; Section~\ref{sec:gsim} summarizes the procedure for extracting from the GARFIELD++ simulation, run once-and-for-all, all the parameters to be used as input for PARSIFAL; Section~\ref{sec:psim} summarizes the simulation procedure inside PARSIFAL and the position reconstruction methods; Section~\ref{sec:tuning} addresses the tuning of the simulation parameters to match the simulated results with the experimental ones; finally the conclusions are drawn.
%
\section{Detector configuration in simulation and experiment} \label{sec:setup}
The detector configuration used in this report is the same as that used in the experimental measurements we performed \cite{jinst} to allow for a direct comparison of the simulated and real data, as described in Section~\ref{sec:tuning}. %
\subsection{Setup and reference frame} 
\begin{figure}[h!]
\centering
\includegraphics[width=0.3\textwidth]{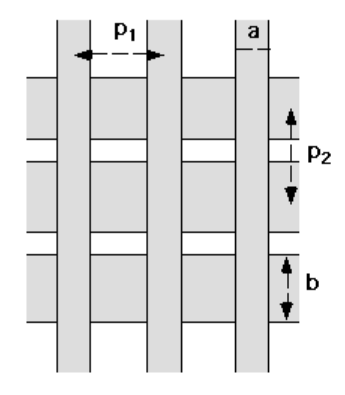} 
\caption{Example of $xy$ readout pattern \cite{gdd}: in our case $p_1 = p_2 = 650~\upmu$m, $a = 130~\upmu$m and $b = 580~\upmu$m.}. \label{fig:anode} 
\end{figure}

The geometric dimensions and electrical settings of the triple-GEM are set as parameters in PARSIFAL, so the description can be easily generalized to different geometries and configurations. \\
As shown in the triple-GEM layout in Figure~\ref{fig:triplegem}, {\it left}, four gaps are defined by the positioning of the electrodes:
\begin{itemize}
\item one {\it drift} gap, between the cathode and GEM~1; 
\item two {\it transfer} gaps, between GEM~1 and GEM~2, GEM~2 and GEM~3; 
\item one {\it induction} gap, between GEM~3 and the anode.
\end{itemize}
Each gas gap is delimited by two electrodes. \\
Figure~\ref{fig:triplegem}, {\it right}, defines the $xyz$ reference frame throughout the paper: the $xy$ plane coincides with the anode plane, while the $z$ axis is orthogonal to it, with positive direction from the cathode to the anode. When no magnetic field is present, the $z$ direction coincides with the direction of the drifting electrons. When the magnetic field is present, it is set orthogonal to the $xz$ plane and its presence affects only the $x$ coordinate, but not the $y$ coordinate. The anode of the experimental setup has two readout planes (double-view readout), segmented in strips with orthogonal directions $x$ and $y$ (see Figure~\ref{fig:anode}). The gas mixture used was $Ar + 10\%\,iC_{4}H_{10}$ at NTP. The APV-25 ASIC \cite{apv25} was used: it samples the signal $27$ times, once every $25$~ns and provides the amplitude of the shaped signal using an ADC. The shaping time is $\tau = 50$~ns.
\subsection{Experimental conditions}
The experimental tests were conducted on the H4 line of SPS in CERN North Area, within the RD51 collaboration \cite{rd51}. A beam of muons with $150$~GeV/{\it c} momentum was available, and this particle type and momentum have been set also in the simulations. Experimental data were acquired both in the absence of magnetic field and with a $1$~T dipole field, with different incident angles between the muons and the GEM chambers, in the range $[0,~40]$ degrees. \\
The experimental data were analyzed and the results can be found in \cite{jinst}. 
\subsection{Summary of the configuration} 
Table \ref{tab:parameters} lists the geometrical and electrical configurations used to characterize the setup and the corresponding values set in the simulations with PARSIFAL, in order to validate them with real data. After validation, PARSIFAL can be extended to other configurations in terms of gas mixtures, geometrical and electrical settings, magnetic fields as well as different types of particles and energies. PARSIFAL was initially implemented for the simplest case of a single-view anode and no magnetic field and later extended to the double-view and magnetic field case, in Section~\ref{sec:2D}. 
\begin{table}[!tb]
\centering
\begin{tabular}{c|l}
\hline
parameter & value \\
\hline
gap thickness                                            & \multirow{2}{*}{$5/2/2/2$ mm} \\
{\footnotesize (drift/transfer~1/transfer~2/induction) } & \\
electric field                                           & \multirow{2}{*}{$1.5/2.75/2.75/5$ kV/cm} \\
{\footnotesize (drift/transfer~1/transfer~2/induction) } & \\
high voltage                                             & \multirow{2}{*}{$275/275/275$ V} \\
{\footnotesize (GEM~1/GEM~2/GEM~3) }                     & \\
gas mixture                                              & $Ar + 10\%\,iC_{4}H_{10}$ \\
strip width/pitch                                        & \multirow{2}{*}{$580/650\,\upmu$m} \\
{\footnotesize ($x$ strip) }                             & \\
strip width/pitch                                        & \multirow{2}{*}{$130/650\,\upmu$m} \\
{\footnotesize ($y$ strip) }                             & \\
distance between $x$ and $y$ strip planes                & $50\,\upmu$m \\
\hline
\end{tabular}
\caption{Geometrical dimensions and electrical settings for the simulated triple-GEM: they are the same as the ones of the setup used in the testbeam. The $x$ and $y$ strips are orthogonal.} \label{tab:parameters}
\end{table}
%
\section{Parametrization of the PARSIFAL input from GARFIELD++ simulations}
\label{sec:gsim}
PARSIFAL is not a competitor of GARFIELD++ but a complementary tool: the variables used in PARSIFAL are extracted from GARFIELD++ simulations, as presented in the following subsections. The simulation consists of five independent parts, which describe ionization, diffusion, amplification, induction and electronics. The first three parts use the preliminary simulations of GARFIELD++, while the induction and electronics are implemented in PARSIFAL from scratch. The assumption that the different steps of avalanche formation and development are independent is an approximation that has been proven valid in a low rate environment \cite{bonivento}. For each set of parameters, a specific simulation must be performed with GARFIELD++, as described below. A summary of the processes to be parametrized from GARFIELD++ to initialize PARSIFAL is listed in Table~\ref{tab:summary}.
\begin{table}[b!]
\centering
\begin{tabular}{r|cl}
\hline
simulation step             & physics process  $\rightarrow$  parameter              & depends on \\
\hline
\multirow{2}{*}{Ionization (Section~\ref{sec:ioni})} & {\small number of primaries/cm}       & gas \\
                             & {\small number of secodaries/primary} & gas \\
\cline{2-3}
\multirow{4}{*}{Drift of electrons (Section~\ref{sec:drift})} & {\small transverse diffusion    $\rightarrow$   $\sigma$ position} & gas, E, B $\neq 0$ \\
                                   & {\small transverse diffusion    $\rightarrow$   $\sigma$ time}    & gas, E, B $\neq 0$ \\
                                   & {\small  longitudinal diffusion  $\rightarrow$   $\upmu$ time}     & gas, E, B $\neq 0$ \\
                                   & {\small  Lorentz force           $\rightarrow$   $\upmu$ position} & E $\times$ B \\
\cline{2-3}
GEM properties (Section~\ref{sec:gemprop})    & {\small  avalanche multiplication $\rightarrow$   effective gain}  & HV \\
\hline
\end{tabular}
\caption{Summary of the physics processes in each simulation step, together with the parameters used to describe them in PARSIFAL (obtained from GARFIELD++ simulations) and with their dependence. E is the electric field, B is the magnetic field (not always present), $\sigma$ and $\upmu$ are the spread and shift in the distributions, HV means high voltage.} \label{tab:summary}
\end{table}
\subsection{Ionization} \label{sec:ioni}
\begin{figure}[h!]
\centering
\includegraphics[width=0.45\textwidth]{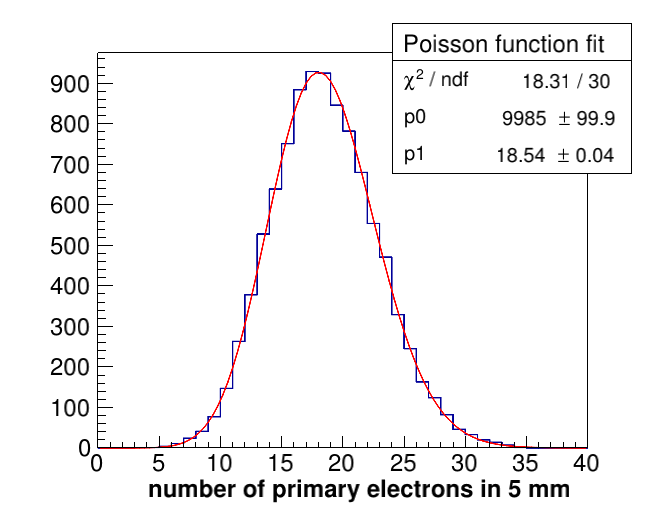} 
\caption{Distribution of the number of primary electrons from ionization in $5$~mm of $Ar + 10\% iC_4H_{10}$, simulated by PARSIFAL. The histogram is fitted with a Poissonian function; the parameters are respectively the normalization constant ($p_0$) and the mean ($p_1$) of the Poisson distribution.} \label{fig:ioniz} 
\end{figure}

Gas-based detectors exploit ionization to detect the passage of particles: the interaction of the charged radiation with the gas atoms generates primary electrons which, if they carry enough energy, further ionize the atoms creating a cluster of secondary electrons. The number of primary electrons from ionization follows a Poissonian statistics \cite{blum}, thus their generation can be simulated by extracting the inter-cluster distances from a proper exponential function \cite{sauli3}. The latter process was studied using GARFIELD++ simulations and the number of secondary electrons in each cluster was tabulated up to $100\,e^-$/cluster. \\
GARFIELD++ simulations were performed: ten thousand muons with a momentum of $150$~GeV/{\it c} were shot into $5$~mm of the simulated gas mixture. The mean number of primary ionizations per centimeter was extracted and used as an input parameter in PARSIFAL along with the number of secondary ionizations for each cluster. \\
The distribution of the number of primary ionizations, resulting from the simulation procedure, is shown in Figure~\ref{fig:ioniz} and can be fitted with a Poissonian function, as expected.
\subsection{Drift of electrons} \label{sec:drift}
\begin{figure}[b!]
\centering
\includegraphics[width=0.49\textwidth]{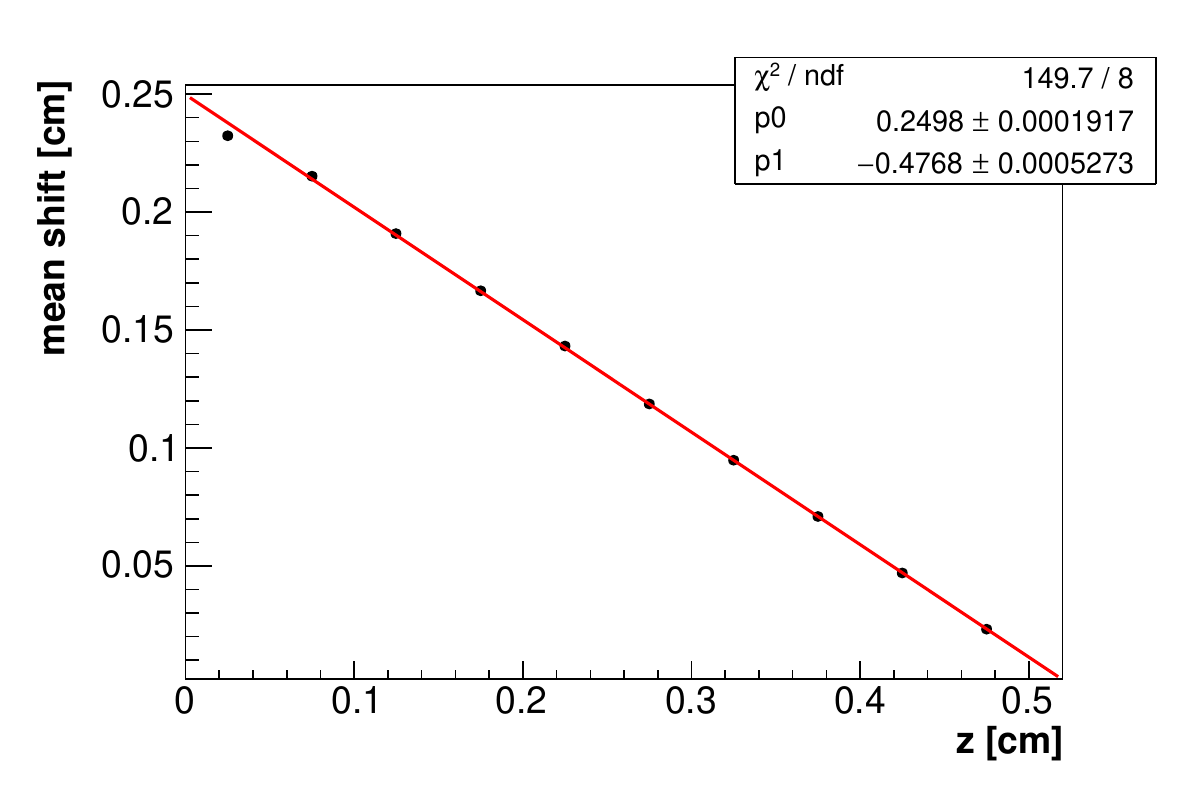}
\includegraphics[width=0.49\textwidth]{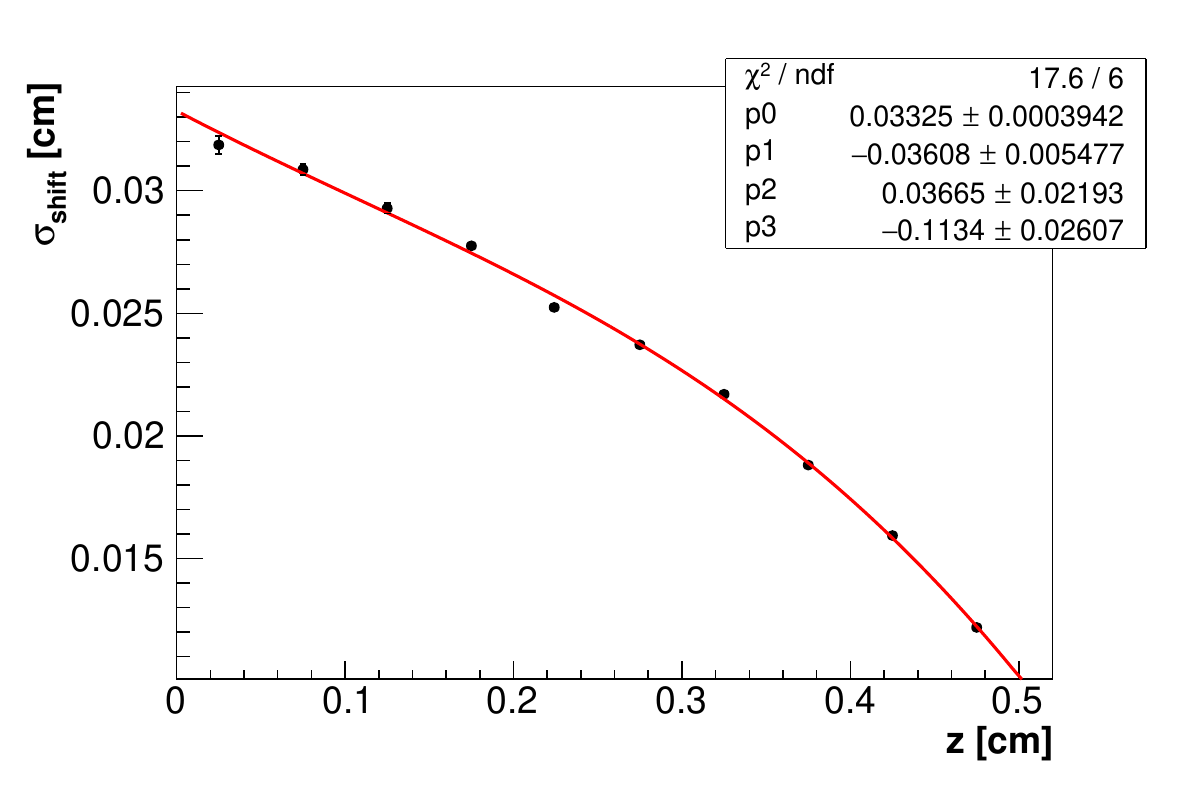}
\caption{Position of the electrons shift ({\it left}) and spread ({\it right}) {\it vs} initial position in the drift gap, from $z = 0$, close to the cathode, to $z = 0.5$~cm, close to GEM~1. The dependence on $z$ is due to the drift path from the different generation points of the ionization electrons. The simulations were run with a magnetic field of $1$~T. The fit is performed with a first degree polynomial $p_0 + p_1 \cdot z$ for the mean shift and a third degree polynomial $p_0 + p_1 \cdot z + p_2 \cdot z^2 + p_3 \cdot z^3$ for the $\sigma_{shift}$. } \label{fig:diff_drift}
\end{figure}

\begin{figure}[h!]
\centering
\includegraphics[width=0.46\textwidth]{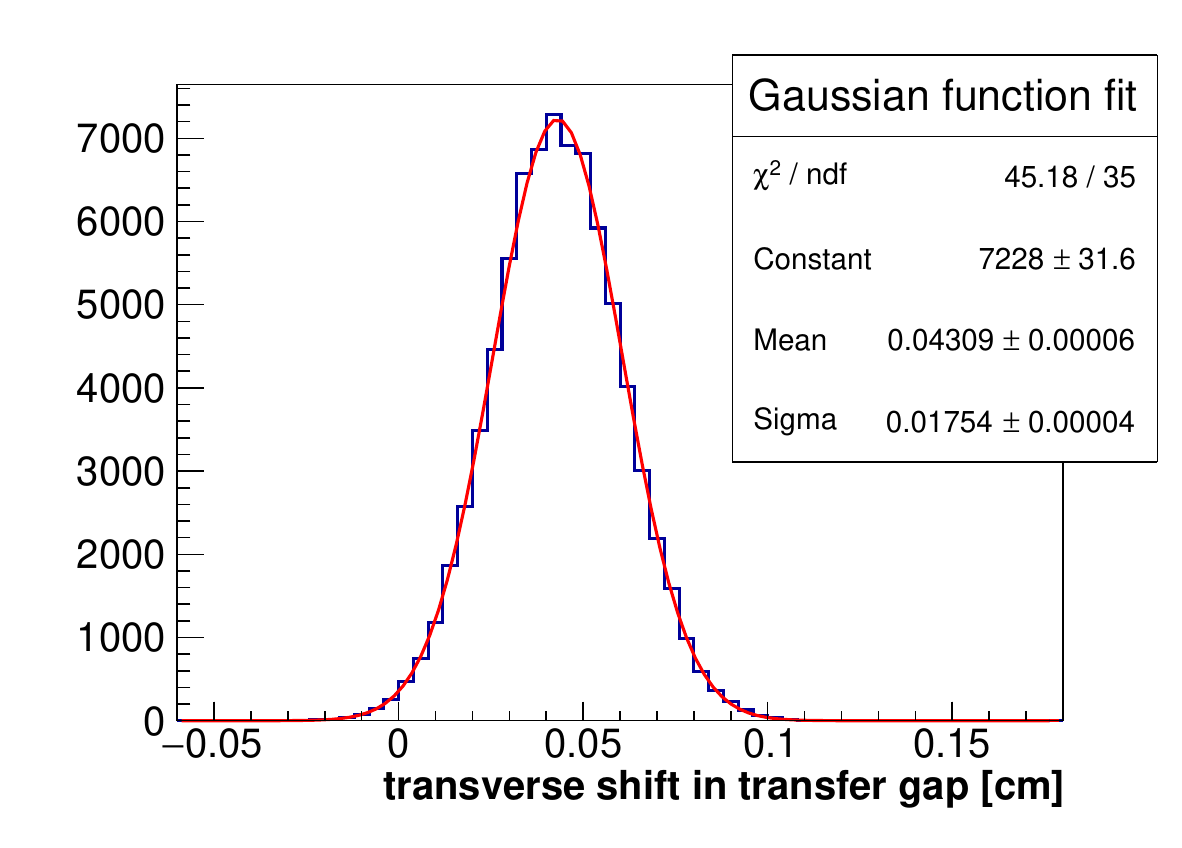} 
\qquad
\includegraphics[width=0.46\textwidth]{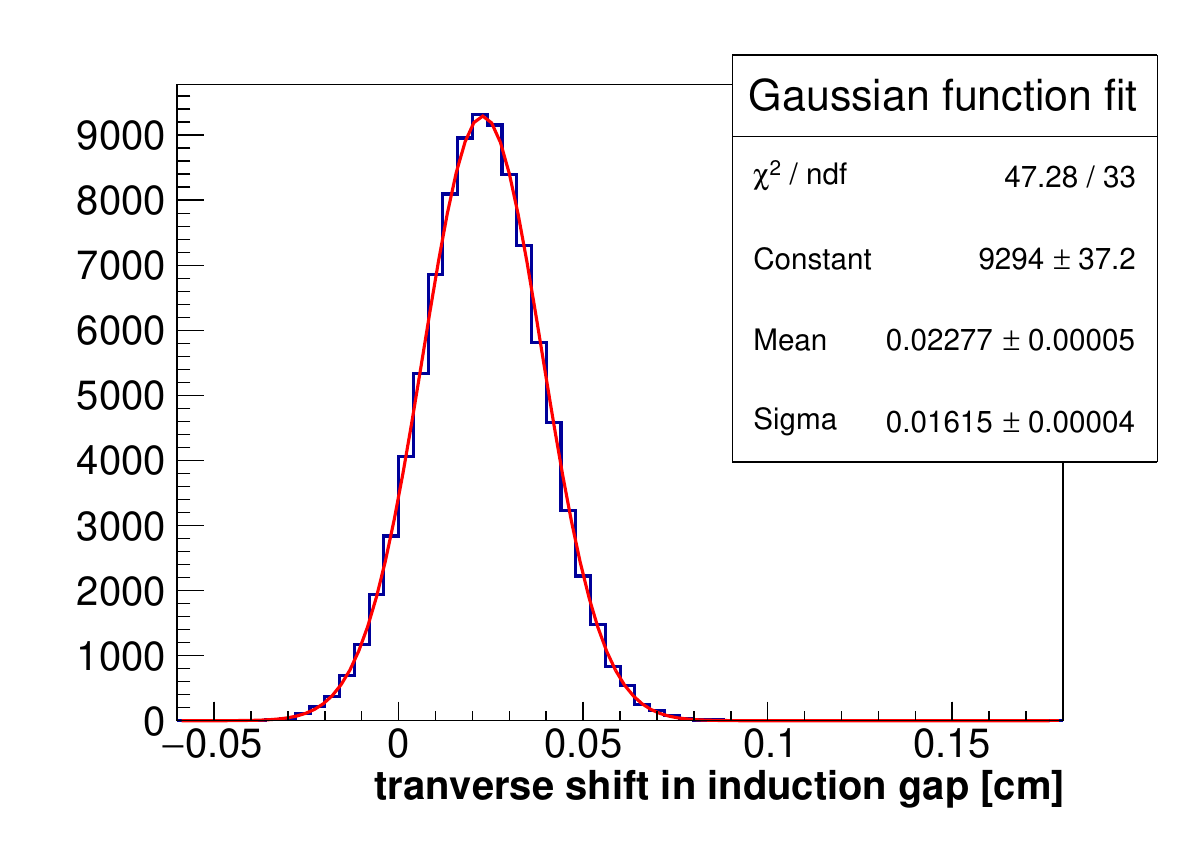} 
\caption{Displacement of electrons on the final plane with respect to the initial position on the starting plane, for the transfer gap ({\it left}) and induction gap ({\it right}). The distributions are Gaussian shaped and not centered in zero since a $1$~T magnetic field is applied. The sigma values of the distributions account for the transverse diffusion effect.} \label{fig:diff_gauss} 
\end{figure}

The electrons from the initial ionization in the drift gap and later the electrons from the avalanche multiplication follow the electric field lines from one electrode to the next one, crossing the different gas gaps. Several factors must be considered in the transport of electrons in a gas under the action of an electric field: the nature of the gas itself, the attachment coefficient, the transverse and longitudinal diffusion, the electric field strength, the electron drift velocity, the Townsend coefficient and the Penning transfer rate. Moreover, the possible presence of a magnetic field modifies the electron motion and must be taken into account in the simulation. \begin{figure}[b!]
\centering
\includegraphics[height=0.37\textwidth]{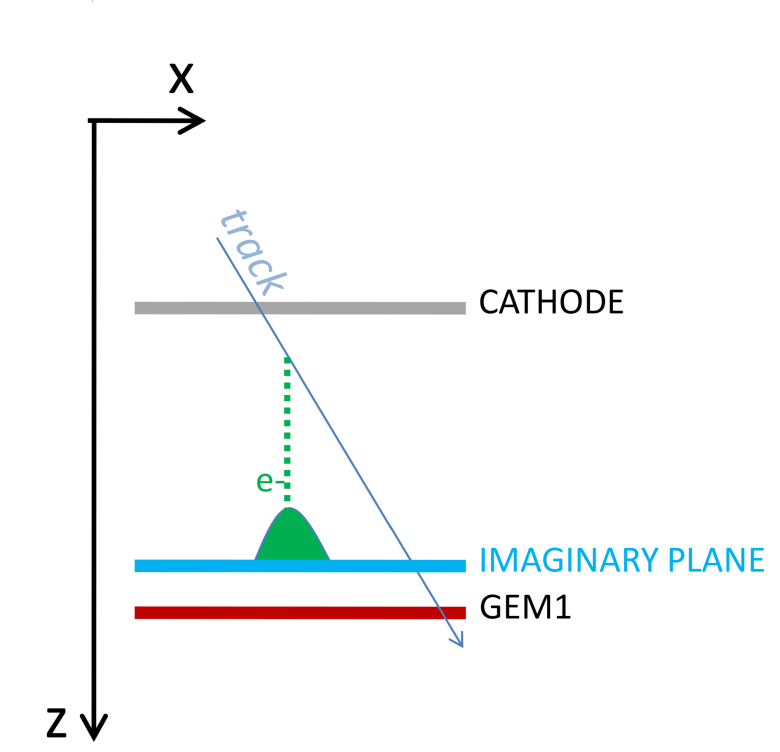} 
\vspace{1mm}
\includegraphics[height=0.37\textwidth]{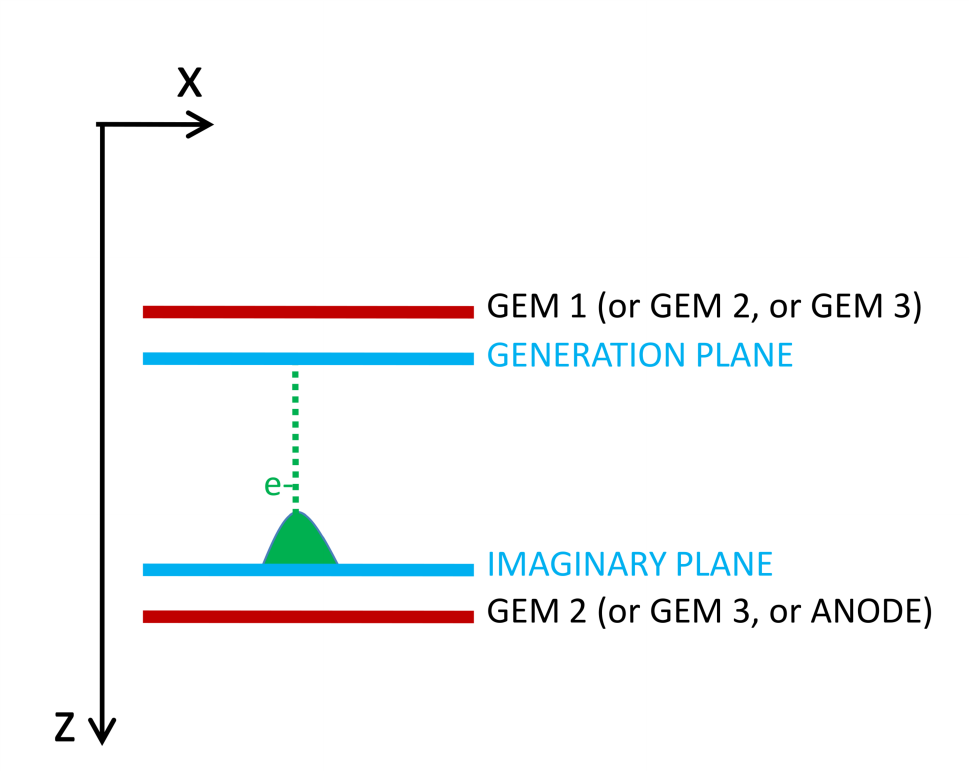} 
\caption{({\it left}) Sketch of the transverse diffusion in the drift gap: the electrons (green dotted line), produced by ionization due to the passage of a charged track (cyan arrow) along its whole path in the drift gap, drift toward GEM~1 and undergo diffusion (represented as a green Gaussian on the imaginary plane). The imaginary plane where this effect is evaluated is $150~\upmu$m before GEM~1. ({\it right}) Sketch of the transverse diffusion in the transfer or induction gaps: here all the electrons enter the gap from the holes of the GEM foil placed before the gap and drift up to the imaginary plane, placed again $150~\upmu$m before the electrode ({\it i.e.} from GEM~1 to GEM~2, from GEM~2 to GEM~3 or from GEM~3 to the anode).} \label{fig:frame}
\end{figure}
%
Despite this large number of factors, the electron position after each gap can be parametrized to speed up the code, by knowing the mean displacement and the position spread as a function of the initial position of the electron: these can be extracted from GARFIELD++. In practice, the effect of the diffusion was evaluated on an imaginary plane placed in the gap region where the electric field lines are still parallel, {\it i.e.} $150\,\upmu$m away from the GEM foils, since by getting closer to the foil the presence of the holes begins to distort the field.
For each gap, ten thousand electrons were simulated and transported through the gap by GARFIELD++ using previously generated electric field maps in the gas mixture. The distribution of the arrival positions on the imaginary plane with respect to the initial position was plotted to extract mean displacement and spread. \\ 
The treatment of the diffusion is the same in all the gaps, except for the drift gap. In the drift gap, electrons are generated from primary ionization along the entire track of the charged particle, hence at different distances from the cathode (see Figure~\ref{fig:frame}, {\it left}). Differently, in the transfer and induction gaps, all the electrons enter the gap from the previous GEM plane, so they all enter the gap at the same distance from the second electrode limiting it (see Figure~\ref{fig:frame}, {\it right}). Therefore, in the drift gap, the final position of the electron depends on its initial $z$, while in the other gaps all electrons have the same initial $z$ coordinate, so this dependence disappears. To take this into account in GARFIELD++ simulations, the individual electrons in the drift gap were generated uniformly in the $z$ direction along the whole gap and transported from the cathode up to the imaginary plane placed $150\,\upmu$m before GEM~1. The drift gap was sliced in several intervals along $z$ axis; for each slice, the position distribution of the electrons at the end of the interval was considered and fitted, to evaluate the mean and sigma values. The trend of the mean {\it vs} distance (as well as the sigma {\it vs} distance) was fitted with a polynomial function and the parameters were plugged into PARSIFAL (see Figure~\ref{fig:diff_drift}).
For the simulation of the two identical transfer gaps, the electrons were generated isotropically on a plane placed $150\,\upmu$m downstream of the GEM foil and transported up to the final plane, placed $150\,\upmu$m upstream of the next GEM (from GEM~1 to GEM~2 and from GEM~2 to GEM~3, see Figure~\ref{fig:frame}, {\it right}). The distribution on the final plane is shown in Figure~\ref{fig:diff_gauss}, {\it left}: it has been fitted with a Gaussian function and the mean and sigma values have been used as input parameters of PARSIFAL. \\
An analogous simulation was performed for the induction gap: the electron motion from a plane $150\,\upmu$m after GEM~3 to the anode was considered. The resulting distribution is shown in Figure~\ref{fig:diff_gauss}, {\it right}.
\\
Despite the ionization occurs along the whole track path, only electrons generated by ionization in the drift gap are considered. Further ionizations in the other gaps are a second order contribution to the signal, since they are multiplied only by one or two GEM stages, rather than three, thus they have been neglected for now. \\
Similarly to the spatial diffusion, the distributions of the drift time of the electrons were studied and inserted among the PARSIFAL input parameters.\\ 
All the simulations were run both with a magnetic field of $1$~T and without magnetic field. Figures \ref{fig:diff_drift} and \ref{fig:diff_gauss} are related to the case with magnetic field, as they both present a shift of the mean value of the distribution under the effect of the Lorentz force. In the case without the magnetic field, the transverse diffusion is still present, but all the distributions are centered around zero.
%
\subsection{Detector effective gain} 
\label{sec:gemprop}
\begin{figure}[tbh]
\centering
\includegraphics[width=0.7\textwidth]{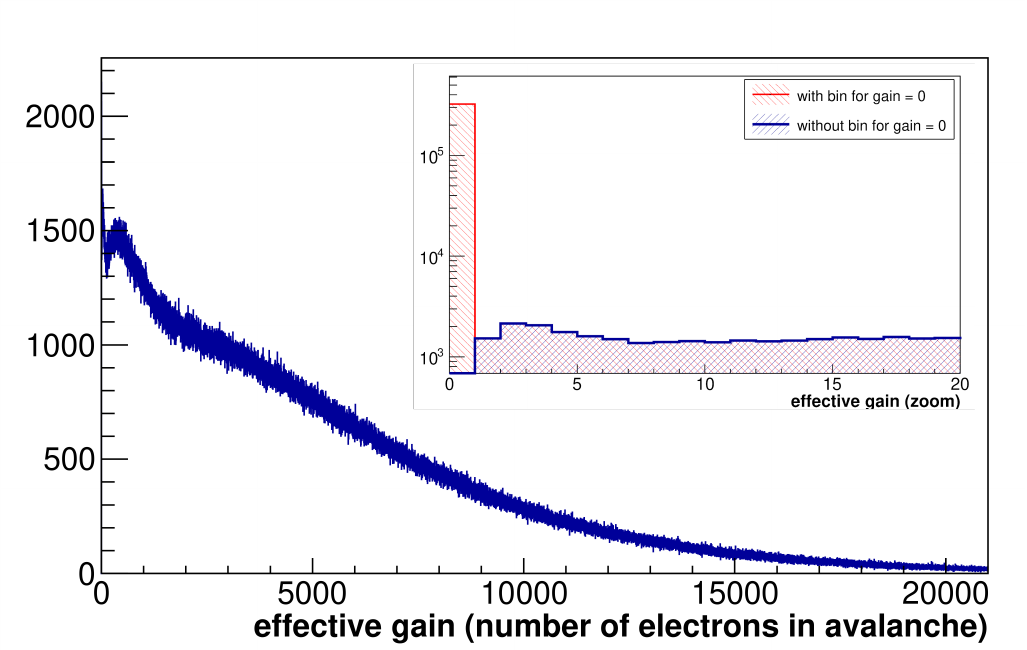} 
\caption{Effective gain for a triple-GEM amplifying component. The peak corresponding to an effective gain equal to zero, due to collection inefficiency of GEM~1, is removed for a better reading of the histogram. The inset shows a zoom of the histogram in the region of gain $<$20: the red histogram shows also the peak for gain equal to zero, to highlight its contribution, about $3\%$ of the entries (note that the $y$ axis is in logarithmic scale). As described in the text, the histogram is the result of the complete simulation of the gain, collection and extraction efficiencies for the three GEMs.} \label{fig:gain} 
\end{figure}

A single foil of GEM is fully characterized by its {\it transparency} and its {\it gain}. \\
In order to obtain the highest achievable gain, all the electrons should be collected into the GEM holes, where the multiplication occurs. Some of the electric field lines which should drive the electrons to the holes, however, fall on the top copper surface of the GEM, hence not all electrons drifting to the GEM contribute to the avalanche. The {\it collection efficiency} $\epsilon_{coll}$ is defined as the number of electrons entering the hole of a GEM, divided by the number of electrons arriving on the GEM plane. We can assume that almost all electrons entering the hole undergo multiplication, but not all of the electrons in the avalanche leave the GEM. Again, some of the electric field lines drive some electrons on the surface of the GEM itself or on the sides of the hole. The {\it extraction efficiency} $\epsilon_{extr}$ is defined as the number of electrons leaving the GEM divided by the number of electrons present in the avalanche. Both efficiencies depend on the electrical settings, in particular, the collection efficiency depends mainly on the field of the gap {\it before} the GEM, while the extraction efficiency depends mainly on the field of the gap {\it after} the GEM. The GEM transparency is defined as the product between the collection and extraction efficiencies. \\
The {\it intrinsic gain} of a GEM is the number of electrons generated in the avalanche per single electron entering the hole. Also an {\it effective gain} can be defined, by multiplying the intrinsic gain by the transparency. The effective gain roughly gives the final number of electrons that can be expected after the multiplication stage, given the initial number of electrons before the GEM. The gain value has an exponential dependence on the high voltage applied to the GEM \cite{bachmann2}. \\
To extract the input parameters for PARSIFAL pertaining GEM behavior, GARFIELD++ simulations were performed separately for each GEM. Ten thousand electrons have been simulated with GARFIELD++ in a region around the GEM, from a plane placed $150\,\upmu$m before it to another placed $150\,\upmu$m after, turning on the avalanche formation; transparencies and gain values have been extracted from these simulations. Variations in the gain of a single GEM are described by the Polya distribution \cite{polya}. 
The parameters extracted from GARFIELD++ were used to evaluate the effective gain of the full triple-GEM, which is not merely the product of the gains of the three GEMs. The effective gain was stored in a histogram (Figure~\ref{fig:gain}) and saved in a file: it is used as input to PARSIFAL, where the effective gain in the simulation is sampled from it.
A warning must be given: GARFIELD++ underestimates the gain of a (single) GEM and the simulated value is a factor two lower than the real one. An explanation of this behavior is still missing and the MPGD community is addressing it \cite{RD51_tuning}. In PARSIFAL, this discrepancy was solved by a tuning factor, which will be described later.
\section{PARSIFAL simulation and position reconstruction} \label{sec:psim}
PARSIFAL reproduces a triple-GEM detector output considering the five physics processes involved independently. The first three, {\it i.e.} ionization, diffusion and multiplication are parametrized starting from GARFIELD++ simulation, as explained in Section~\ref{sec:gsim}; the last two, {\it i.e.} induction and electronics, are described in this Section. \\
The ionization, diffusion and multiplication provide the spatial and temporal distributions of the charge on the readout as follows:
\begin{itemize}
\item from ionization, the number of primary electrons is sampled and their positions are generated;
\item from multiplication, the effective gain associated with each primary electron is sampled;
\item from diffusion, the final position of each electron in the avalanche is sampled.
\end{itemize}
%
In order to obtain results comparable to the experimental data, a description of the response of the APV-25 ASIC is included in the code, but it can be replaced by other readout electronics if needed. Moreover, the reconstruction of the particle position has been implemented.
\subsection{Induction} \label{sec:induct}
\begin{figure}[h!]
\centering~
\includegraphics[width=0.47\textwidth]{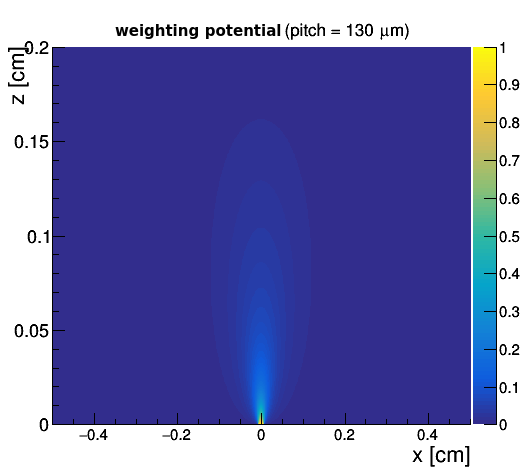}
\includegraphics[width=0.47\textwidth]{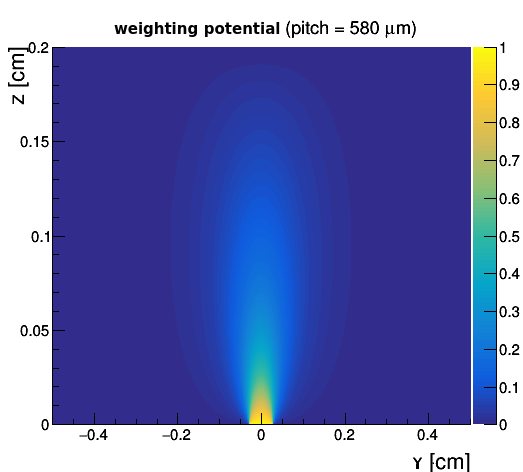}
\caption{Analytical calculations of the weighting potential, for a strip size of $130\,\upmu$m ({\it left}) and $580~\upmu$m ({\it right}). The strip is centered in $(0,0)$ and an induction gap of $2$~mm is considered, the color scale unit is the volt.} \label{fig:wf}
\end{figure}

The signal in the triple-GEM detector is read from the strips on the anode plane, which are connected to the electronics. The motion of the electrons in the induction gap induces a time-dependent current on the strips, continuously from the moment they exit GEM~3 until their arrival on the anode. Once {\it all} the electrons have arrived on the anode, the signal is over. The instantaneous current $i_{ind}(t)$ induced on the $i_{th}$-strip is given by the Shockley-Ramo theorem \cite{shockley,ramo} as:
\begin{equation}
i_{ind}(t) = {\it e} \cdot \mathbf{v}_{drift}(t) \cdot \mathbf{E}_w(\mathbf{x}(t)) \label{eq:ramo},
\end{equation}
where {\it e} is the electron charge, $\mathbf{v}_{drift}$ is the drift velocity, $\mathbf{E}_w$ is the {\it weighting field} and $\mathbf{x}(t)$ is the electron position at the time $t$. The weighting field is a computational artifice and corresponds to the electric field generated by the electrode under consideration, when kept at $1$~V, with all the other electrodes set to $0$~V. The analytical calculation of its $z$ component $E_{1z}$ (from \cite{rieglerwf}) is
\begin{equation} \label{eq:riegler}
E_{1z} = -V_1 \frac{1}{2 D} \bigg[ \frac{ \sinh{(\pi \frac{x-w/2}{D})}}{ \cosh{(\pi \frac{x-w/2}{D})} - \cos{(\frac{z \pi}{D}})} - \frac{ \sinh{(\pi \frac{x+w/2}{D}})}{ \cosh{(\pi \frac{x+w/2}{D})} - \cos{(\frac{z \pi}{D}})} \bigg],
\end{equation}
where $V_1=1$~V, $x$ and $z$ are the coordinates of the point where the calculation is performed, on and orthogonal to the anode respectively, in a reference frame whose origin coincides with the strip position, $D$ is the induction gap thickness and $w$ is the strip width. This calculation, however, does not consider the non-active area between the strips (pitch $\equiv$ active area) and it does consider only one strip plane\footnote{The strips in the simulation are parallel to the $y$ direction, being the anode plane segmented along the $x$ axis.}: correction factors are needed to account for both approximations. Moreover, in PARSIFAL the drift velocity of the electrons towards the anode is simply assumed to be constant and set to the value extracted from GARFIELD++ simulations. In addition, the weighting potential was used instead of the field and the calculation of the induced current from Equation~(\ref{eq:ramo}) was replaced by the computation of the charge $Q_{ind}(dt)$ induced in the infinitesimal time interval $dt = t_2 - t_1$ in which the electron moves from the point $\mathbf{x}(t_1)$ to $\mathbf{x}(t_2)$ with a difference in the weighting potential equal to $\Delta{V_w}$:
\begin{equation}
Q_{ind}(dt) = {\it e} \cdot \Delta{V_w} \cdot \cos\phi \sim {\it e} \cdot \Delta{V_w},
\label{eq:ramo2}
\end{equation}
where $\Delta{V_w} = V(\mathbf{x}(t_1)) - V(\mathbf{x}(t_2))$ is the gradient of weighting potential and $\phi$ is the angle between the weighting field and the drift velocity. The weighting potentials computed analytically for strip sizes of $130~\upmu$m and $580~\upmu$m (dimensions used in our setup) are shown in Figure~\ref{fig:wf}.
Two simulations of the induction process were implemented: a full and a fast induction. \\
{\bf Full induction} - PARSIFAL simulation computes the final position on the anode for each electron in the avalanche which enters the induction gap, by applying the diffusion in the gas and the Lorentz angle displacement. In order to build the electronic signal and its correct time development, the induced charge $Q_{ind}(dt)$ is computed for each time step $dt = 1$~ns, on all the strips around the final point of arrival of the electron. The contributions to the induced charge from all the electrons are summed up. 
\\
{\bf Fast induction} - The electronic signal ends once all the electrons have been collected by the strips and there is no charge carrier moving in the induction gap anymore. During the movement of the electrons from the entrance point in the gap towards the anode, positive and negative currents can be induced on the strips, but once all the electrons have been collected, the total charge induced on one single strip corresponds to the total number of electrons actually falling on that strip. This is also clear from Equation~(\ref{eq:ramo2}) if we consider that at the point of entrance in the induction gap the $V_w=0$ while on the collecting electrode it is $V_w=1$V, hence $Q_{ind}=e$. For this reason, the fast induction was added to PARSIFAL, to obtain a reliable signal using a much faster method. Here, each electron entering the induction gap is extrapolated to a strip of arrival. This takes into account the electron starting position at the entrance of the induction gap, the magnetic field and the diffusion effect. Each electron contributes to the strip charge with one electronic charge, released at the time the electron arrives at the anode. \\
The full induction provides the charge induced on the $i_{th}$-strip, while the fast induction provides the charge collected on it. The proper agreement of the induced and collected charge values was verified, by simulating the induction of the same electron with both methods. The fast induction provides a huge improvement in computing time, as it is around $30$ times faster than the full induction. 
\subsection{Double-view readout} \label{sec:2D}
\begin{figure}[h!]
\centering
\includegraphics[width=0.7\textwidth]{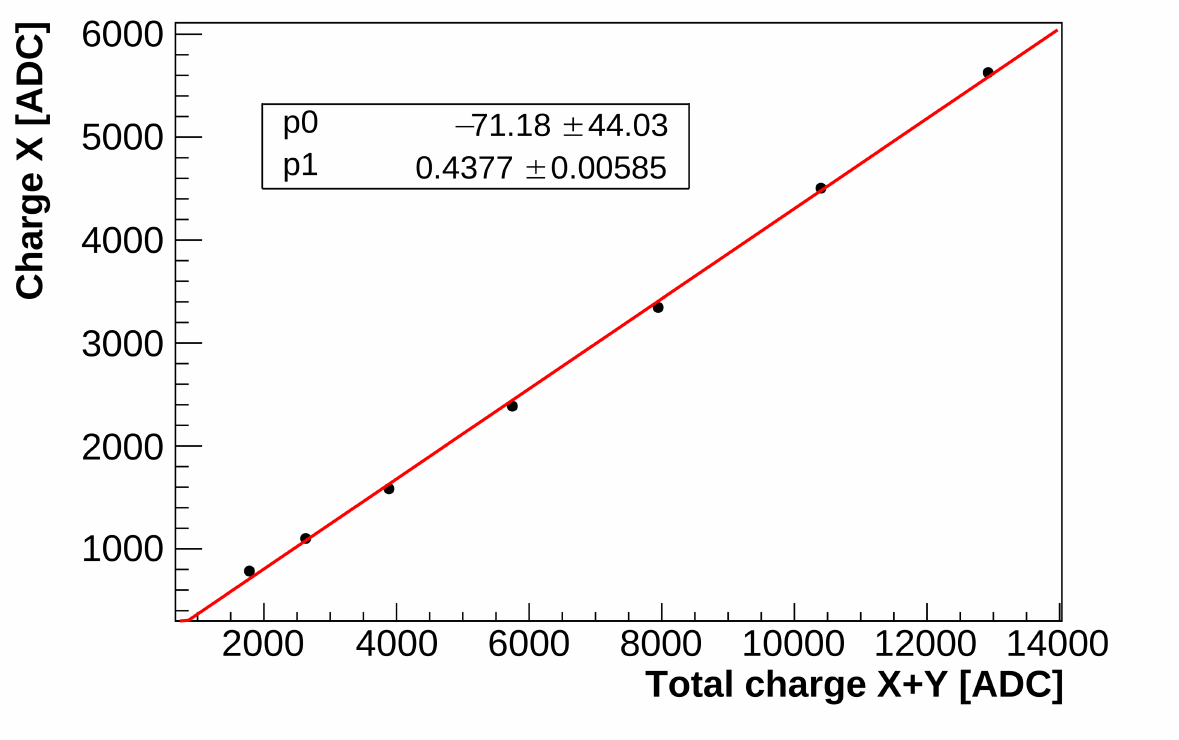}
\caption{Charge collected on view $x$ {\it w.r.t} total charge in a high voltage scan, on experimental data. The charge sharing (CS) results as $\frac{Q_y}{Q_x} = 1.3$.} \label{fig:sharing}
\end{figure}

The anode used to collect the experimental data has a double-view readout, so two planes of strips collect the induced charge. 
In PARSIFAL, the total amount of electrons is split between the views: one electron is assumed to be collected only by one strip and for large numbers the charge sharing is granted. Several charge measurements, acquired at different gain values, have been used to evaluate the charge sharing between $x$ and $y$ strips from the experimental data. In Figure~\ref{fig:sharing}, the charge on the $x$ view is plotted as a function of the total charge and a linear fit extracts the ratio between the charge collected by this view and the total charge. Based on the experimental data, the charge sharing results $\frac{Q_y}{Q_x} = 1.3$.
%
\subsection{Electronics} \label{sec:ele}
\begin{figure}[h!]
\centering
\includegraphics[width=\textwidth]{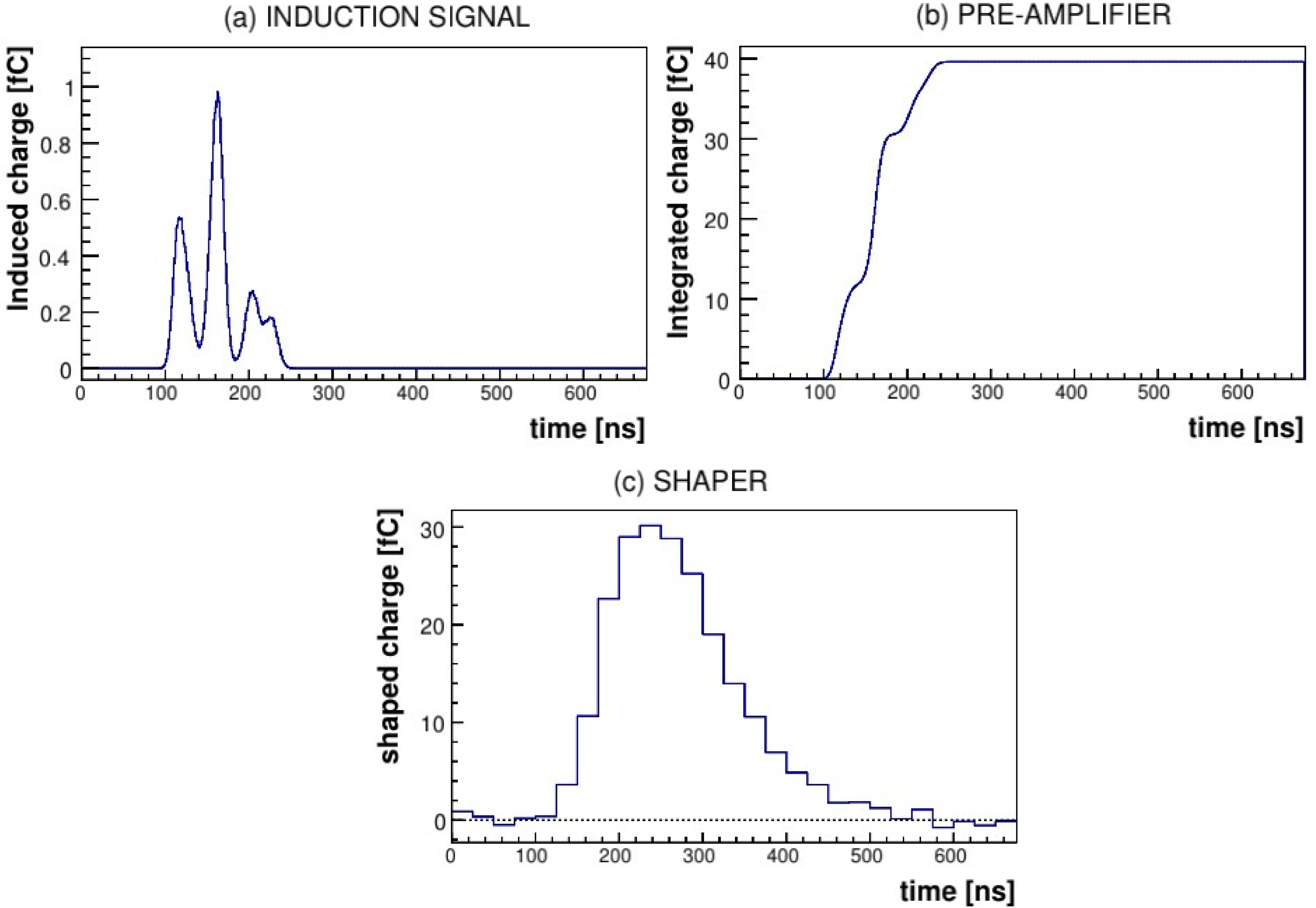}
\caption{Simulation of the induction and electronics readout: induced charge on a strip (a), pre-amplifier output (b), shaper output (c).} \label{fig:apv25}
\end{figure}

Since the electronics plays a role in the measurement, a simplified simulation of the same ASIC used in the testbeam, the APV-25, was implemented in the code. Each APV-25 channel was simplified as a pre-amplifier and a shaper \cite{apv25}. The pre-amplifier was modeled as a simple integrator, without the implementation of any electronic gain. It integrates the signal {\it continuously}, so our approximation provides the integration every $1$~ns (see Figure~\ref{fig:apv25}, (b)). The shaper is simplified as a CR-RC circuit, with shaping time $\tau = 50$~ns. In order to compute the {\it shaped} integrated charge $Q_{shaped} (t)$ for each APV-25 time bin ($27$ time bins, each $25$~ns long), the following shaping function is applied:
\begin{equation} \nonumber
Q_{shaped} (t) = Q_{preamp} \, (\frac{t - t_0}{\tau}) \, \exp{(-\frac{t - t_0}{\tau})}, \label{eq:shaping}
\end{equation}
with $Q_{preamp}$ as the integrated charge from the pre-amplifier, $t$ as time bin time, $t_0$ as time of beginning of the signal. For each time bin $t_i$, the electronic signal $Q_{\textrm{\scriptsize APV-25}}$ is the the sum of all the shaping functions coming from previous time bins, evaluated at the time $t_i$: 
\begin{equation} \nonumber
Q_{\textrm{\scriptsize APV-25}} (t_i) = \sum_{j \leq i}{Q_{shaped, j}(t_i)}.
\end{equation}
For each strip and each time bin, a random noise is added, sampling a Gaussian function centered in zero and with a sigma evaluated from the data (see Figure~\ref{fig:apv25}, (c)).
%
\subsection{Reconstruction of data} \label{sec:reco}
\begin{figure}[tb!]
\centering
\includegraphics[width=0.6\textwidth]{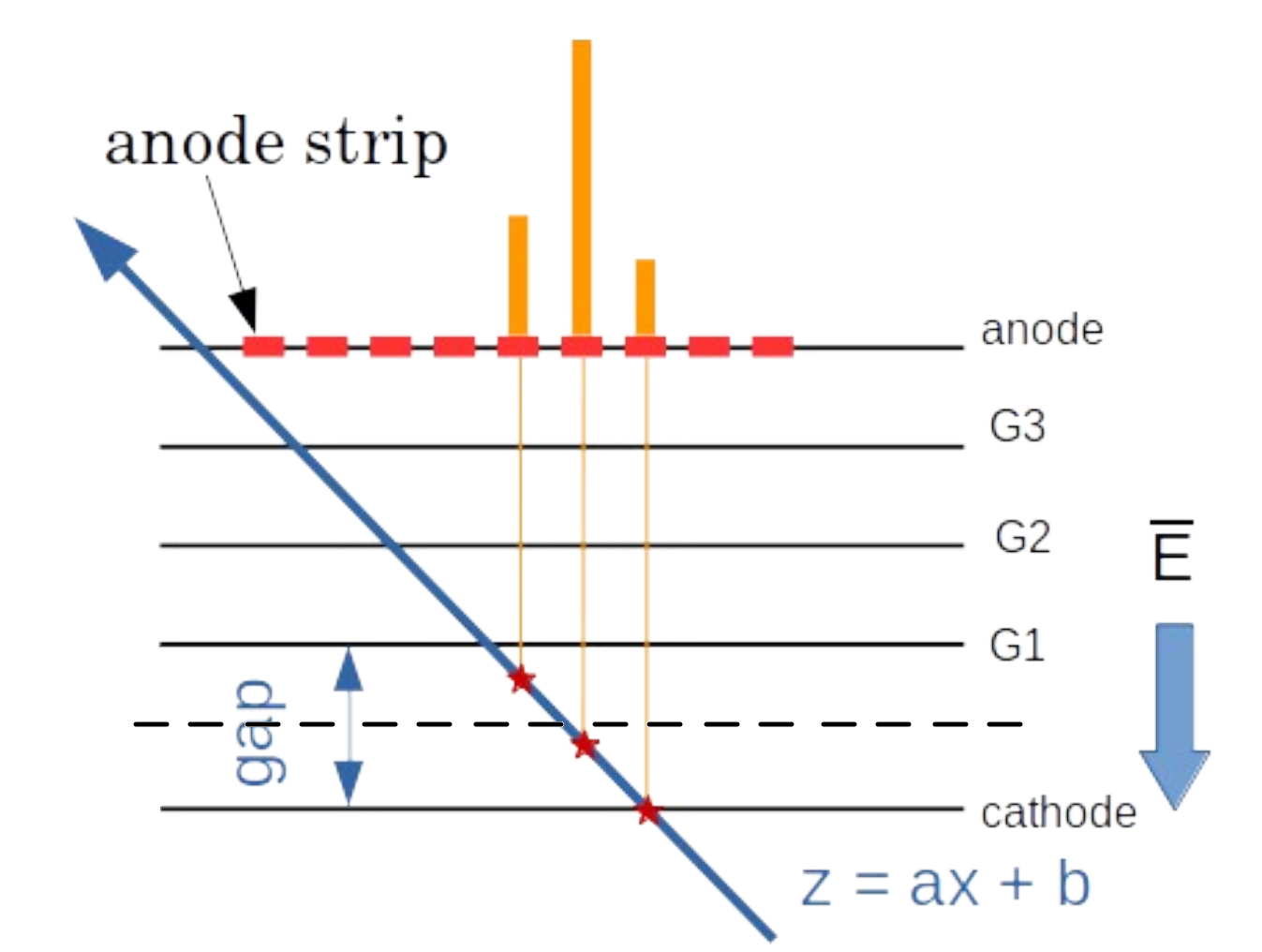} 
\caption{Description of the {$\upmu$}TPC method. The primary ionization positions are represented by the red stars, the electron drift by the orange lines and the charge collected on the strips by the orange bars. A linear fit is applied to the points and the slope $a$ and the constant $b$ are used in the $x_{{\upmu}TPC}$ method. The position is reconstructed in the middle of the drift gap, highlighted by the dashed line. In the depicted case there is no magnetic field.} \label{fig:utpc}
\end{figure}

The simulated signal resembles with good accuracy the real one: the rise time is about $120$~ns in both cases. The {\it hit charge}, {\it i.e.} the charge measured on each strip, comes directly from the maximum value of the $Q_{\textrm{\scriptsize APV-25}}$ histogram (see Figure~\ref{fig:apv25}, (c)), while the {\it hit time} comes from fitting the rising edge of the signal with a Fermi-Dirac function \cite{jinst}. The simulated and experimental charge will be compared in Section~\ref{sec:tuning}, where the simulation is tuned to the experimental data and the code is finally validated. \\
Thanks to the good agreement of the simulated signal with the experimental one, both can be reconstructed in the same way. A strip is fired if the measured charge is above a certain threshold $Q_{thr}$. When more adjacent strips are fired, they can be grouped to form a {\it cluster}. The total cluster charge as well as the multiplicity of strips in one cluster, defined as {\it cluster size}, are two of the variables used to evaluate the goodness of the simulation when compared to real data (see Section~\ref{sec:tuning}). \\
The position can be reconstructed via two algorithms: the charge centroid (CC) and the micro-Time Projection Chamber ({$\upmu$}TPC) mode. Hence, the other two important quantities to evaluate the simulation are the positions reconstructed via CC and via {$\upmu$}TPC methods. \\
The CC method computes the cluster position $x_{CC}$ as the average of the positions $x_i$ of the fired strips, weighted by the charge $q_i$ measured on each of them, as follows: 
\begin{equation} 
x_{CC} = \frac{\sum_{i = 0}^{\textrm{\scriptsize cl.size}}{x_i\,q_i}}{\sum_{i = 0}^{\textrm{\scriptsize cl.size}}{q_i}}. \label{eq:cc}
\end{equation}
The {$\upmu$}TPC method uses the drift gap as a time projection chamber of few millimeters. When considering the $x$ strips, the {$\upmu$}TPC points lie on the $xz$ plane. 
Each {$\upmu$}TPC point ($x_i, z_i$) is described by two coordinates: $x_i$ is the position of each strip, the $z_i$ coordinate is obtained by multiplying the drift velocity, known from GARFIELD++ simulations, and the signal time measured on each strip, from the Fermi-Dirac fit. The pair $(x_i, z_i)$ is the bi-dimensional position of each primary ionization in the drift gap. By fitting all these points with a line $z = a \cdot x + b$ (as shown in Figure~\ref{fig:utpc}), the position $x_\textrm{\scriptsize {$\upmu$}TPC}$ in the middle of the drift gap can be computed as: 
\begin{equation}
x_\textrm{\scriptsize {$\upmu$}TPC} = \frac{{gap}/2 - b}{a} \label{eq:utpc},
\end{equation}
where $gap$ is the drift gap thickness, $a$ is the slope and $b$ the intercept of the fitting line (see Figure~\ref{fig:utpc}).
\\
The same reconstruction algorithms are applied to the $y$ coordinate.
\section{Tuning to experimental data} \label{sec:tuning}
The first requirement for a detector simulation is to replicate the experimental results with a very good accuracy: this makes the validation of the code a necessity to ensure a reliable prediction. Since some approximations are used in the simulation, their impact is discussed in Section~\ref{sec:comments}, before the actual description of the tuning procedure is given in Section~\ref{sec:tunproc}. 
\subsection{General comments on simulation} \label{sec:comments}
PARSIFAL, in its current layout, can fully simulate the passage of a single charged particle much faster than GARFIELD++. 
While the current implementation processes the events sequentially, the possible exploitation of the parallel computing, applied to the drift of electrons and the avalanche formation, could improve in future the CPU-time gain even further. \\
The reduction in computing time was achieved also thanks to the introduction of some approximations:
\begin{itemize}
\item the origin of secondary electrons from ionization in the drift gap is the same, in position and time, as that of primary electrons;
\item the signal is created only by electrons from ionization in the drift gap: in reality, about $2\%$ of the signal is due to electrons created in the first transfer gap, which undergo only two multiplication stages \cite{poli}. For now, this contribution has been neglected;
\item for full induction, the drift velocity is considered as constant ($37\,{\upmu}$m/ns, in drift gap); 
\item for full induction, the weighting field/potential is computed analytically with only one strip plane, while an anode with two orthogonal strip planes is considered in this paper. To cope with this discrepancy, a multiplication factor to account for the charge sharing was introduced;
\item for fast induction, the time of arrival of each electron on the strip is taken as the time associated with each charge deposition, while the induction of the signal is a continuous process throughout the entire drift in the induction gap.
\end{itemize}
\subsection{Tuning procedure without magnetic field} \label{sec:tunproc}
\begin{figure}[h!]
\centering
\includegraphics[width=\textwidth]{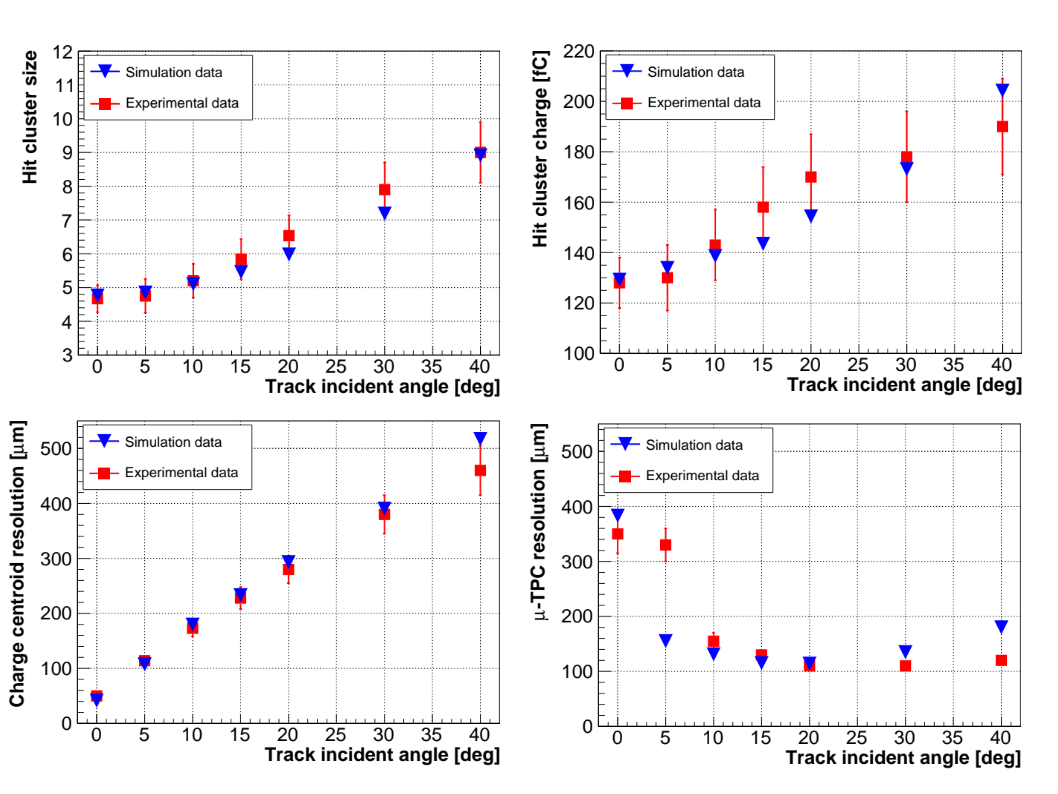}
\caption{Comparison between the simulated ({\it blue triangles}) and experimental ({\it red squares}) data for the four sentinel variables {\it vs} the incident angle: on top the cluster size ({\it left}) and charge ({\it right}), on bottom the spatial resolution for CC ({\it left}) and {$\upmu$}TPC ({\it right}). The experimental data analysis can be found in \cite{jinst,riccardo}. No magnetic field is used in these measurements.} 
\label{fig:tuning1}
\end{figure}

In order to check and tune the results obtained by the simulation, the experimental data collected with the setup described in Section~\ref{sec:setup} were used and four {\it sentinel variables} were chosen as indicators of the goodness of the agreement: the cluster charge, the cluster size, the spatial resolution on the position reconstructed with CC and with {$\upmu$}TPC methods. These four variables were used in \cite{jinst} to characterize the chambers in the testbeam, so their choice here is natural. Charge and cluster size are related to the electrical settings and the gas mixture, so they characterize well the triple-GEM detector behavior when changing the fields and high voltage. The spatial resolution with the CC and {$\upmu$}TPC methods characterize the tracking performance. These are the macroscopic variables that can be measured experimentally and fully describe the GEM behavior, which is described microscopically in GARFIELD++ simulations. The macroscopic variables can be compared to the results from PARSIFAL simulations which are based on GARFIELD++ microscopic description, to evaluate the reliability of the code. \\
A first version of the tuning is focused on two main elements: the gain value and the spatial diffusion. Other parameters are left un-tuned and their value is taken directly from the experimental data. The conversion factor, to convert the ASIC ADC steps to fC, is fixed to the testbeam value, {\it i.e.} $30\,$ADC$\,=\,1\,$fC \cite{kostas}; in the same way, the charge threshold was not tuned ($Q_{thr} = 45$ ADC). The noise level, estimated from random trigger runs to be about $15$ ADC, was not tuned as well. A more detailed version of the tuning procedure is planned for the future to include the fine optimization of all these values. \\
{\bf Gain} - The simulated gain needs to be corrected by a tuning factor for each GEM foil since it is based on GARFIELD++ output, which is underestimated by a factor ranging from $1.5$ to $2.5$ with respect to the experimental gain. This is a known issue of GARFIELD++ \cite{RD51_tuning,factortwo1,factortwo2,factortwo3,factortwo4}.  \\
{\bf Spatial diffusion} - The simulated spatial diffusion needs to be tuned in order to achieve the matching of the resulting cluster size and {$\upmu$}TPC reconstructed position resolutions. In fact the simulated cluster size is lower with respect to the experimental one and this generates also a mismatch in the {$\upmu$}TPC reconstructed position resolution. \\
The tuning procedure consists in spanning over a set of combinations of detector gain and diffusion tuning values, among which the one which best matches the experimental data is selected. The simulations are run for each combination, on a set of seven incident angles $[0,\,5,\,10,\,15,\,20,\,30,\,40]$ degrees and for each parameter $p$, the corresponding $\chi^2_p$ is computed. It is defined as in Equation~(\ref{eq:chi2each})
\begin{equation}
\chi^2_p = \sum_{i=0}^{N_\textrm{\scriptsize angle}}{\frac{
( p_{i}^{sim} - p_{i}^{exp})^2}{\sigma_{p_i}^{exp}}}. 
\label{eq:chi2each}
\end{equation}
Only the error in the experimental data is considered since the statistical error in the simulation is negligible. The samples used for the tuning consist of ten thousand simulated particles for each setting resulting in a statistical error of about $1\%$, while the error in the experimental data was evaluated to be around $10\%$ for the spatial resolution and $5\%$ for cluster size and charge. The simulations for the tuning were run with the fast induction and without magnetic field.\\
To perform the minimization on the four sentinel variables simultaneously, a global $\chi^2$ is computed as the sum of the $\chi^2_p$ on the single parameters: 
\begin{equation}
\chi^2 = \chi^2_Q + \chi^2_{cl.size} + \chi^2_{CC} + \chi^2_{{\upmu}TPC}. \label{eq:chi2}
\end{equation}
In order to select the best parameters, the reduced $\chi^2_R = \chi^2/\nu$ (with $\chi^2$ from Equation~(\ref{eq:chi2}) and $\nu$ the number of degrees of freedom) has been considered: its value must approach one. To determine the tuning factor of the detector gain, the values from $1$ to $30$ were scanned, in steps of $0.1$; for the tuning factor of the diffusion the values from $1$ to $3$ were scanned, in steps of $0.1$. The best combination of parameters scored a reduced $\chi^2_R \sim 3$, for a detector gain tuning factor of $14.1$ and a diffusion tuning factor of $1.5$. \\
The comparison between the simulated and experimental data, with the best tuning settings, is shown in Figure~\ref{fig:tuning1}. The agreement is remarkable for all the angles. The environmental conditions (temperature, pressure and humidity) for the points at $5$ and $15$ degrees differ slightly with respect to the other points and this could explain the not perfect agreement in the plot of the {$\upmu$}TPC resolution. The $5$ degree angle falls in between the region where the {$\upmu$}TPC method performs well (at higher angles) and where its performance is poor (at smaller angles), so the non optimal agreement in correspondence of this incident angle is mostly due to these effects. No magnetic field was used in these measurements, but it will be shown that the found tuning factors can be applied also to the case with magnetic field by obtaining an equally good agreement (see Section~\ref{sec:results}).
\subsection{Validation of the tuning in magnetic field} \label{sec:results}
\begin{figure}[h!]
\centering
\includegraphics[width=\textwidth]{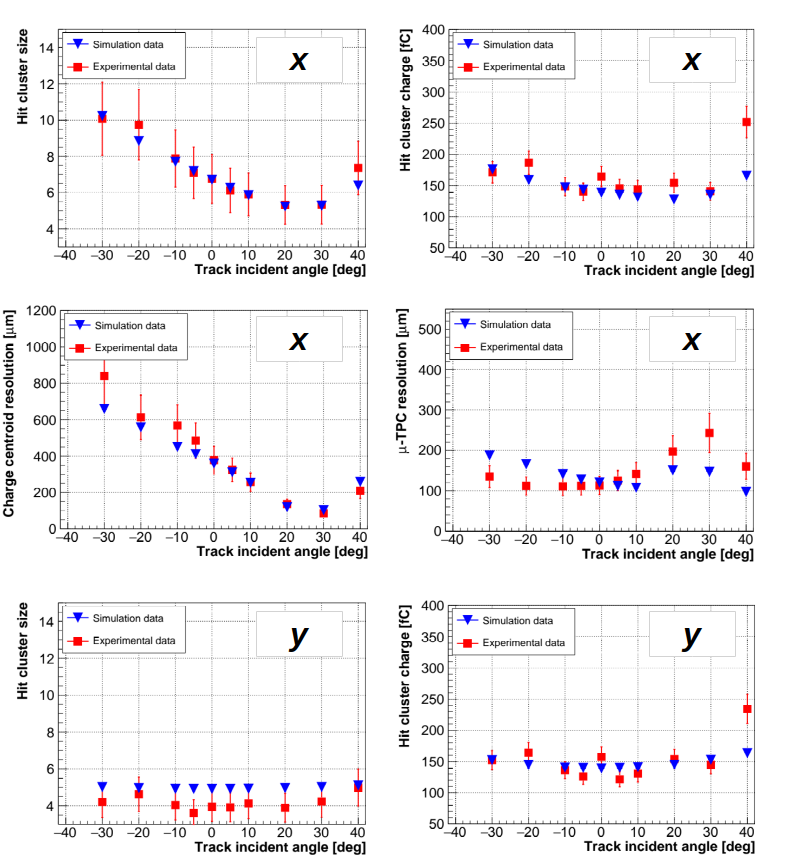}
\caption{Comparison between the simulated ({\it blue triangles}) and experimental ({\it red squares}) data for the sentinel variables {\it vs} the incident angle, in magnetic field of $1$~T and considering the two views. The experimental data analysis can be found in \cite{jinst,riccardo}.} \label{fig:tuning2}
\end{figure}

The tuning parameters were obtained using a set of data collected without a magnetic field. To verify and validate the simulations, in order not to bias the results, they were tested on a different set of data, collected during the same testbeam, with the magnetic field switched on. The matching between the experimental and simulated values for this data is shown in Figure~\ref{fig:tuning2}. Figure~\ref{fig:tuning2} highlights that the tuning factor evaluated without magnetic field grants an optimal matching between experimental and simulated data also with magnetic field. \\
Table \ref{tab:accordance} shows the level of agreement for each of the tuned parameters plotted in Figure~\ref{fig:tuning2}, in terms of number of sigmas ($\mathcal{A}_\sigma$), defined as the ratio
\begin{eqnarray} 
\mathcal{A}_\sigma & = & (V_{exp} - V_{sim})/\sigma_{exp}, \label{eq:loa1}
\end{eqnarray}
where $V_{exp}$ and $\sigma_{exp}$ are the experimental value of a generic variable and its associated error and $V_{sim}$ is its simulated value.
\begin{table}[h!]
\centering
\begin{tabular}{clc}
\hline
view & variable & $\mathcal{A_\sigma}$ \\
\hline
X & cluster size  & 0.17 $\sigma$  \\
X & cluster charge & 1.08 $\sigma$ \\ 
X & CC resolution  & 0.67 $\sigma$ \\
X & {$\upmu$}TPC resolution & 1.36 $\sigma$ \\
Y & cluster size  & 1.00 $\sigma$ \\
Y & cluster charge & 1.02 $\sigma$ \\
\hline
\end{tabular}
\caption{Level of agreement $\mathcal{A_\sigma}$ from Equation~(\ref{eq:loa1}) for each sentinel variable plotted in Fig.\ref{fig:tuning2}.} \label{tab:accordance}
\end{table}
A level of agreement around $1\,\sigma$ shows that PARSIFAL can properly simulate the behavior of a triple-GEM detector. The agreement in the {$\upmu$}TPC reconstruction also shows the correctness of the time simulation, since this method relies heavily on the time measurements. There is room for further improvement, mainly by removing some approximations and refining some simulations, but this agreement further proves the potential of the simulation with PARSIFAL. Of course, this can and must be done by preserving the computing speed, which is the key factor of PARSIFAL implementation. Also, additional tuning factors may be required to adapt the simulation to the experimental data even better than $1\,\sigma$.
\section{Conclusion} \label{sec:future}
A parametrized simulation of a triple-GEM has been successfully implemented: PARSIFAL is able to simulate ionization, drift, gain and induction independently of the detector specific geometry. For this reason, it can be applied also to other MPGDs with simple adjustments. \\
The quality of PARSIFAL results, shown by its validation against testbeam data, proves that it is possible to use it to simulate the full avalanche development and the signal formation in a triple-GEM chamber, in a small amount of time (less than two seconds). The level of agreement with the results obtained on real data is about $1\,\sigma$. \\
The parametrization begins with the detector description in GARFIELD++. The simulated signal is extracted from the ionization, amplification, diffusion and induction; it reproduces the real signal shape thanks to the tuning factors that are taken into account to overcome the limits of the introduced approximations. Among the tuning factors, the one for the gain is particularly interesting: a value of $14.1$ for the detector gain, corresponding to a gain tuning factor of $2.4$ on the single GEM, was found and the results are in agreement with other evaluations in the literature \cite{RD51_tuning}. \\
As stated, PARSIFAL can be extended to various configurations of geometry, high voltage settings and gas mixtures. Moreover, since some of the physics involved is shared among different technologies, the code can be extended to simulate other gaseous detectors. In fact, it can be adapted directly to other MPGDs, {\it e.g.} MicroMegas and $\upmu$-RWELL. \\
The code will be released to the MPGD community repository and on the GIT repository at the link: https://github.com/Hilldar/PARSIFAL, in the branch named triplegem.
\section*{Acknowledgments}
This research was funded by the European Commission in the RISE Project 645664-BESIIICGEM, H2020-MSCA-RISE-2014 and in the RISE Project 872901-FEST, H2020-MSCA-RISE-2019.
\bibliographystyle{elsarticle-num}
\bibliography{parsifal_paper}
\end{document}